\documentstyle[prl,aps,twocolumn,psfig]{revtex}
\begin{document}
\draft\preprint{}
\title{The phase diagram of the lattice Calogero-Sutherland model}
\author{E. Le Goff, L. Barbier and B. Salanon}
\address{CEA/DSM/DRECAM/SRSIM \\
Service de Recherches sur les Surfaces et l'Irradiation de la Mati\`ere \\  CEA/Saclay 91191 Gif Sur Yvette Cedex France}
\date{\today}

\maketitle

\begin{abstract}
We introduce a {\em lattice} version of the Calogero Sutherland model
adapted to describe $1/d^2$ pairwise interacting steps with discrete positions on a vicinal surface. The configurational free energy is obtained within a transfer matrix method. The full phase diagram for attractive and for repulsive interaction is deduced. For attraction, critical temperatures of faceting transitions are found to depend on step density.
 \end{abstract}


 Among one dimensional quantum models the Calogero Sutherland (CS) model has attracted much attention \cite{sutherland70,sutherland71}. In this model fermions on a line interact with a pairwise $1/d^2$ potential. As an important application, the CS model was shown to be equivalent to a model of steps with $1/d^2$ interaction, although in a continuous version, i.e. with no underlying lattice \cite{jayaprakash84}. This approach produced numerous useful results among which the analytic form  of the free energy in the repulsive case \cite{sutherland70,sutherland71,williams93}. This exact result exerted a major influence on subsequent approaches, nonetheless the  CS model is a continuous one, with no crystallographic sites. As a consequence it can exhibit no roughening transition. In addition, the model is ill defined below a certain critical temperature $T_0$ when the interaction is attractive \cite{jayaprakash84}. In this case, although there are mean field indications that faceting occurs at low temperature \cite{jayaprakash84}, no solution for a better behaved model was suggested, nor were lattice effects taken into account when using the CS model in the framework of vicinal surfaces.

In this article, in order to obtain a generic approach to study the instabilities of vicinal surfaces, we introduce a two dimensional model for steps which can be seen as a {\em lattice} CS model \cite{noteHS}. Since the lattice introduces a minimal distance between steps (or fermions) this model is well behaved for all temperatures. For repulsive interaction  it unsurprisingly exhibits the Kosterlitz Thouless (KT) roughening transition. For attractive interaction we find first order faceting transitions which depend on step density. Using the transfer matrix method and finite size scaling we determine the full orientational phase diagram in the repulsive and in the attractive case as well. 

In the field of surface physics the technological applications of nanostructures make the understanding of the thermodynamical stability of surfaces a crucial issue. At a given temperature it is the behavior of the surface free energy as a function of orientation which determines the stability/instability of any surface orientation, as well as the equilibrium crystal shape (ECS) \cite{herring51,andreev81}. On the experimental side the ECS  was determined in several cases \cite{heyraud80,rottman84b,metois87}. Orientational phase diagrams of $Si$, $Pt$ and $Au$ were drawn from x-ray experiments\cite{yoon94,song95,watson98}. On the theoretical side there have been several attempts to describe the energy of stepped surfaces in terms of pairwise interactions between steps. The form of step-step interactions is known in the case of elastic repulsion, which varies as $A/d^2$ \cite{marchenko81}. It is also known when electronic dipoles are located on steps, then the interaction is also $1/d^2$, although with either sign, according to dipole orientation. 
 
  Although complex effects such as terrace reconstruction are likely to play a role in some experimentally observed cases of faceting, by using the CS model, we focus on generic aspects of instabilities.

 Our model depends on two parameters, the kink creation energy $E_K$ and the interaction amplitude $A$. For $N$ steps separated on average by $l_0$ sites and with periodic boundary conditions the configurational energy for a periodic cell is written as :

\begin{eqnarray}
{\cal H}=\sum_{m,y} |h_{m,y+1}-h_{m,y}|E_K+\nonumber\\ 
\label{H}
\sum_{m'>m,y} V(h_{m',y}-h_{m,y})+{\pi^2 \over 6} {A \over N l_0^2}\\
V(d)=\sum_{n=-\infty}^\infty{A \over (d+nL)^2}=A{\pi^2 \over L^2}\left[\sin({\pi d \over L}) \right]^{-2}\nonumber
\end{eqnarray}

where $m$ and $m'$ vary from $1$ to $N$, $h_{m,y}$ is the position of step $m$ at site $y$ along the step direction. $L=Nl_0$ is the size of the periodic cell.

 The first term of $\cal H$ in Eq.(\ref{H}) is the contribution of kinks. The next two terms reflect pairwise step step $A/d^2$ interactions. The potential $V$ was introduced in order to include interactions between steps belonging to different periodic cells, in addition to the interaction between steps belonging to a given cell \cite{sutherland70}. The third term is the interaction of steps in the cell of interest with their periodic images, this term is absent in $V$. When taking the Boltzmann constant equal to unity the $g$ parameter of the CS model is $g(T)=2A/Tb^2(T)$, where $b^2(T)=(\cosh(E_K/T)-1)^{-1}$ is the $T$ dependent step diffusivity. 

We determine here the orientational phase diagram of the model i.e. critical temperatures in the $(p,T)$ plane. The step density $p=1/l_0$ varies between $0$ and $1$. In the following we take integer values for $l_0$.

In order to localize transitions we perform a Finite Size Scaling (FSS) analysis  of the free energy obtained with the Transfer Matrix (TM) method on a cylinder (see e.g. \cite{luck83,dennijs92}). We thus treat $N$ (with $N$ from 2 to 6) infinitely long steps with transverse periodic boundary conditions, the free energy depend on the largest eigenvalue of the TM. Our goal in using FSS is twofold. On one hand, from the theory of conformal invariance, the variation of the free energy with $N$ is a criterion for critical states \cite{affleck86}. On the other hand, the behavior of the $(N\rightarrow\infty)$ extrapolated free energy is a criterion for the thermodynamic stability of phases.  
 For the FSS analysis of rough states it is of interest to introduce a coarse grained energy as a Fourier expansion on wave vectors ${\mathbf q}=(q_m,q_y)$ \cite{villain85}:

\begin{equation}
{\cal F}=\sum_{q_mq_y}\left[ \eta_m\left(1-\cos q_m \right)+\eta_y\left(1-\cos q_y \right) \right]|h_q|^2
\label{calF}
\end{equation}

where $\eta_m$ (resp. $\eta_y$) is the stiffness in the $m$ (resp. $y$) direction. For this Gaussian model the FSS properties can be readily determined and the free energy {\em per terrace site} can be written :

\begin{equation}
F(N)=F_\infty-p{\pi T \over 6} \sqrt{\eta_m \over \eta_y}{1 \over N^2} \label{F(N)}
\end{equation}

In the perspective of conformal field theory the central charge $c$ is unity, since the system is rough in the sense of KT. The sound velocity $v$ is thus directly related to the anisotropy in the $m$ and $y$ directions by $v=p\sqrt{\eta_m/\eta_y}$ (in agreement with \cite{legoff99}).  

Then, for a given $p$, we examine the convergence of the free energy obtained with the TM method by fitting numerical values with :

\begin{equation}
F(N)=F_\infty-{a\over N^\alpha} 
\label{Fit}
\end{equation}

\begin{figure}
\psfig{figure=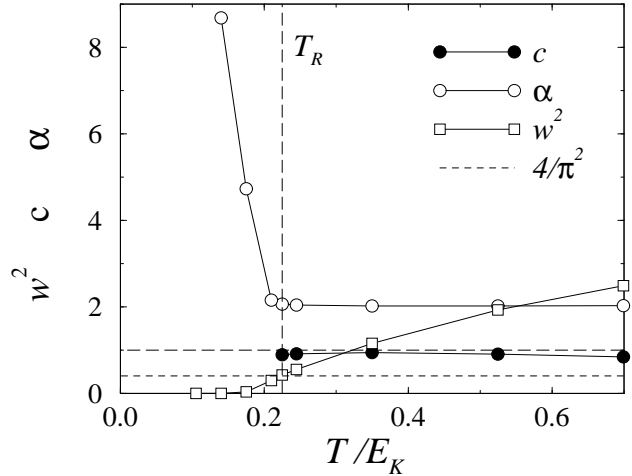,width=8cm}
\caption{For $p=0.25$ the free energy of the model of Eq.~(\ref{calF}) for a finite cylinder is used to produce successive values of $c(N)$, the plotted values of $c$ are power law extrapolations. The values of $\alpha$ are obtained by fitting data with Eq.~(\ref{Fit}). Within the capillary approximation $w^2$ is equal to $4/\pi^2$ at $T_R$.}
\label{alpha}
\end{figure}

\begin{figure}
\psfig{figure=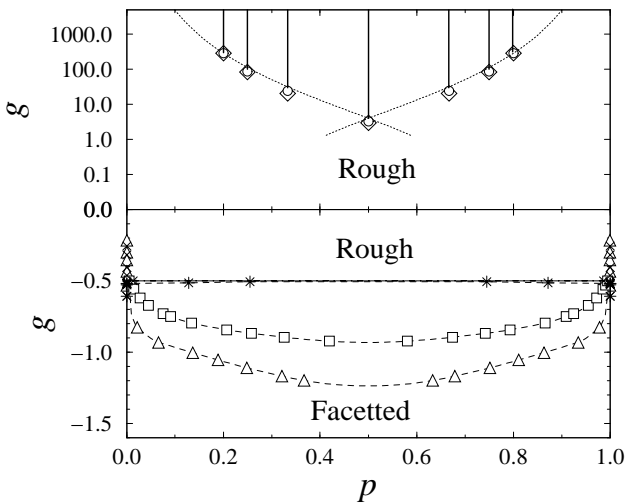,width=8cm}
\caption{Orientational phase diagram. Repulsion ($g>0$) :  $A/E_K=0.014$ ($\circ$), $A/E_K=0.140$ ($\diamond$). Vertical lines indicate the flat regions. Dotted lines represent the criterion for $T_R$ as in Eq.~(\ref{crittr}) for steps and for holes. Attraction ($g<0$) : $T_0/E_K=0.20$ ($\triangle$), $0.60$ ($\Box$) and $1.00$ ($\ast$). Dashed lines are guides for the eye.}
\label{diag_phase}
\end{figure}

For $g>0$, rough states are critical and $\alpha$ is expected to be equal to $2$. Fig.~(\ref{alpha}) shows the values of the apparent exponent $\alpha$ resulting from the fit for $p=0.25$. In the region where $\alpha=2$ the system is rough. When using the coarse grained energy of Eq.~(\ref{calF}) for a finite cylinder with approximate values of $\eta_m$ and $\eta_y$ \cite{legoff99,barbier96} to estimate $v$, we find for the apparent central charge $c=0.92\pm0.02$. With exact values of $\eta_m$ and $\eta_y$ the real central charge would thus be nothing but unity. In contrast, when the apparent value of $\alpha$ is greater than 2 the system is flat. The roughening temperature is located at the threshold. This can be confirmed by other types of arguments.  Thus we have calculated with the TM method the standard deviation of the terrace width distribution $w^2$, which takes the special value ${4 \over \pi^2}$ at the roughening temperature $T_R$ \cite{legoff99}. $T_R$ can also be determined by solving \cite{villain85,legoff99}:

\begin{equation}
2g(T_R)=(p^{-2}-1)^2 -1.
\label{crittr}
\end{equation}
 
  which is in principle valid for small $p$. It is apparent on Fig.~(\ref{alpha}) that all criteria lead to very close values of $T_R$ and we have made the phase diagram of Fig.~(\ref{diag_phase}) with the $w^2$ criterion.

Note that in the fermionic approach \cite{sutherland70} and subsequent papers, the interaction of steps with their periodic images in other cells was missing (third term in Eq.~(\ref{H})). This has led to two different determinations of the central charge. $c=1$ was believed to be the exact value \cite{kawakami91,romer92} while from \cite{sutherland70} one gets $c=\lambda$ with $\lambda=(1+\sqrt{1+2g})/2$. When the missing term ${N\over L}{\pi^2\over 6}{g \over L^2}={\pi^2 \over 3}{\lambda(\lambda-1) \over l_0L^2}$ is included the fermion energy per site becomes $E_f={1 \over 3}{\lambda^2\pi^2 \over l_0^3}-{1 \over 3}{\lambda\pi^2 \over l_0^3}{1 \over N^2}$. As the sound velocity is $v={2\pi \lambda \over l_0}$, this shows that the central charge is indeed unity. Neglecting this term would have made the FSS analysis inconsistent.

  As step-step interactions are pairwise one can invoke the step-hole symmetry of the Hamiltonian in Eq.~(\ref{H}) to show that the phase diagram exhibits a $p\rightarrow1-p$ symmetry. The criterion in Eq.~(\ref{crittr}) can thus be used for holes and the phase diagram is symmetric with respect to $p=1/2$ . One sees that Eq.~(\ref{crittr}) gives an acceptable estimation for $T_R$ even for $p=1/2$, where it crosses the corresponding criterion for holes. Nonetheless it turns out that  $g(T_R)$ indeed varies when varying $A/E_K$, so that the scaling of Eq.~(\ref{crittr}) is weakly broken by lattice effects around $p=1/2$. 

In order for the free energy to reflect the step-hole symmetry and exhibit more readily its convexity/concavity properties we use $F_s(p)=F(p)-pF(1)$ with $F(1)={\pi^2\over 6}A$. It is then convenient to introduce a symmetric Landau like expansion of $F_s$. It turns out that excellent fits of $F_s(p)$ are obtained with the following form :

\begin{equation}
\phi(p)=b_0+b_2(p-{1 \over 2})^2+b_4(p-{1 \over 2})^4+b_6(p-{1 \over 2})^6
\label{phi}
\end{equation}

 with the constraints that $\phi(0)=0$ and that the slope for $p=0$ is $\beta-{\pi^2\over 6}A$, with $\beta$ the free energy per unit length for an isolated step. The coefficients $b_i$ are temperature dependent. In the rough phase (low $p$) we find a good agreement between extrapolated values of the free energy $F(p)$ and the  classical analytical form \cite{sutherland70,sutherland71,jayaprakash84,williams93}:

\begin{equation}
\gamma(p)=f_0+{\beta}p+{\delta}p^3
\label{f}
\end{equation}

where $f_0$, the surface energy of the reference plane, is here taken equal to zero and $\delta(T)$ is given by \cite{sutherland71,williams93}:
 
\begin{equation}
\delta(T)={\pi^2\over 6}Tb^2(T)\lambda^2 \text{~with~} \lambda={1 \over 2} \left(1+\sqrt{1+2g}\right)
\label{delta}
\end{equation}

Note that when $l_0$ is not an integer the roughening temperature can be much lower than the value given by Eq.~(\ref{crittr}) \cite{schultz85}.
  
In the attractive case ($g<0$) it is known that fermions collapse for $g< -1/2$. It was conjectured that a tricritical point would appear and that faceting would occur if the potential divergency at the origin were cut off \cite{jayaprakash84}. In our {\em lattice} model steps cannot collapse and the orientational phase diagram can be investigated. There are two parameters, the kink creation energy $E_K$ and the attraction strength $A$, or equivalently the temperature $T_0$ defined by $g(T_0)=-1/2$.
We have made calculations for $T_0/E_K=0.20$, $0.60$ and $1.00$, in order to see the influence of this ratio.

\begin{figure}
\psfig{figure=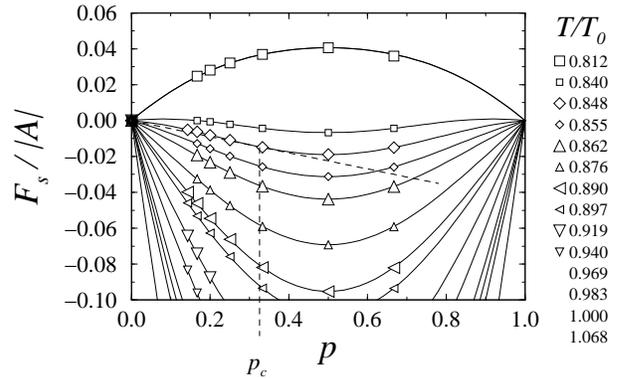,width=8cm}
\caption{Symmetric free energy $F_s(p)$ (see text) as a function of step density $p$, $T_0/E_K=0.60$. TM results extrapolated with Eq.~(\ref{F(N)})(symbols), fit with expansion as in Eq.~(\ref{phi}) (------), tie line for $T/T_0=0.848$ (-- -- --). $T_1/T_0=0.838$.}
\label{F(p)1}
\end{figure}

We find that for $T>T_0$ one has $\alpha=2$ for all tested values of $p$. The free energy is convex, the entropic repulsion is dominant, so that all orientations are stable and rough. For $T<T_0$ one finds that $\alpha$ increases above $2$ when $T$ crosses a $p$ dependent temperature $T_c(p)$. Correlatively the orientational form of the free energy  changes and $F_s(p)$ becomes concave around $p=0$ and $p=1$ (see Fig.~\ref{F(p)1}). This indicates that the related orientations $[0,p_c]$ and $[1-p_c,1]$ are unstable with respect to faceting. Of course, within the thermodynamical limit ($N\rightarrow\infty$) TM calculations would give directly rise to phase separation and the free energy would converge like $1/N$ due to the presence of walls. Except at very low $T$, this is not the case and it is clear that concave regions are stabilized by finite size effects. The values of the corresponding free energy is presumably influenced by such effects whereas the tie line is not, very much alike convex regions. From the fitted $p-1/2$ power expansion of $F_s(p)$ we determine the boundaries of the two-phase (faceted) $p$ range. For $T_0/E_K=0.20$ and $T_0/E_K=0.60$ the unstable range increases with decreasing $T$. Below a critical temperature $T_1$, $F_s(p)$ becomes concave in the whole $p$ range and faceting occurs between $p=0$ and $p=1$. Fig.(\ref{F(p)1}) shows extrapolated values of $F_s(p)$ for various temperatures in the vicinity of $T_0$. 

\begin{figure}
\psfig{figure=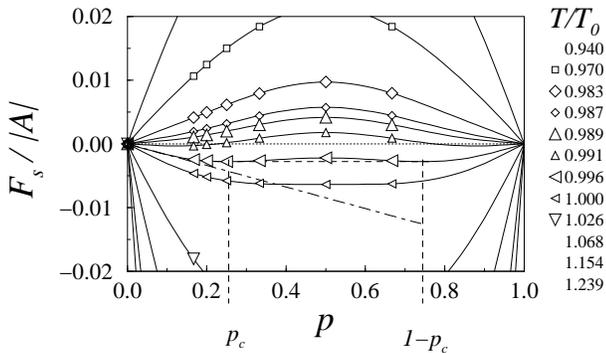,width=8cm}
\caption{Same as Fig.~(\ref{F(p)1}) for $T_0/E_K=1.00$. Tie lines for $T/T_0=0.996$ (--- -- --- --) and (-- -- --). $T_1/T_0=0.989$.}
\label{F(p)2}
\end{figure}

It is noticeable that, for $T_0/E_K=1.00$, $T_1$ is very close to $T_0$. Only in the very narrow range of temperature $[T_1,T_0]$ does faceting occur close to $p=0$ and $p=1$. Moreover, as $b_2$ turns out to be negative for $T_1<T<T_0$, an unstable $p$ range appears around $p=1/2$, as shown on Fig.~(\ref{F(p)2}).  Results for the three cases are summarized on Fig.~(\ref{diag_phase}).

As theoretical predictions and the analysis of experimental results rather use the expansion of $F$ in powers of $p$ as  $F(p)=f_0+a_1|p|+a_2p^2+a_3|p|^3+...$ 
it is of interest to show the coefficients $a_i$ deduced from our $b_i$. It is noticeable that, the higher $T_0/E_K$, the narrower the $p$ range where $F_s(p)$ is correctly given by the first four terms of the $p$ expansion. 

\begin{figure}
\psfig{figure=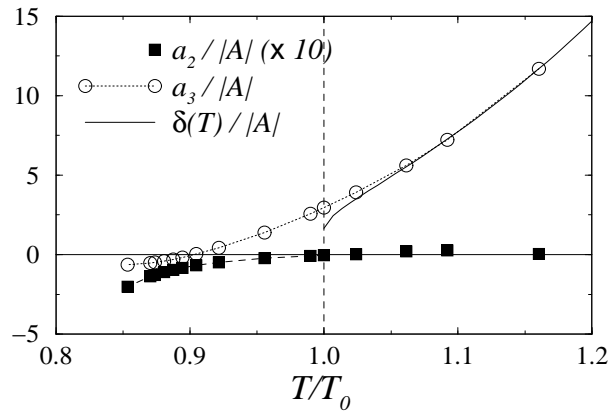,width=8cm}
\caption{Coefficients of the power expansion of $F_s(p)$: $a_2$, $a_3$ and  $\delta(T)$ as in Eq.~(\ref{delta}) as functions of $T$, dashed and dotted lines are guides for the eye, $T_0/E_K=0.20$.}
\label{ai}
\end{figure}

For $T_0/E_K=0.20$ this limited expansion is valid for $p<0.5$. Fig.~(\ref{ai}) shows the values of $a_2$ and $a_3$ as functions of $T$. Above $T_0$ the value of $a_3$ is in agreement with $\delta(T)$ as given by Eq.~(\ref{delta}) whilst around $T_0$ lattice effects bring about strong deviations. Thus $a_3$ remains positive slightly below $T_0$, then it becomes negative, which indicates that the metastable phase computed with the TM method is uniform in this $T$ range. The coefficient $a_2$, which fluctuates around zero above $T_0$, becomes negative below $T_0$, thus driving the instability and showing that the tricritical point is very close to $T_0$. The presence of a quadratic term for $T<T_0$ is in contrast with mean field (Hartree Fock) approaches which give no such term \cite{jayaprakash84}.  For $T_0/E_K=0.60$ and $1.00$ the full power expansion should be used for $p>0.1$ in the whole $T$ range.

We have thus shown that lattice effects alter the main features of the continuous CS model. In the repulsive case, while the previously known roughening transition is recovered, we show evidence for the symmetry related roughening of holes. In the attractive case the {\em lattice} CS model is well behaved and the critical temperature for first order faceting becomes $p$  (step density) dependent. For vanishing $p$ (and $1-p$) the critical temperature tends to $T_0$, the limiting temperature in the {\em continuous} CS model. In the ($p,T$) phase diagram ($0,T_0$) and ($1,T_0$) appear as tricritical points. This phase diagram should be of interest for all lattice realizations of the CS model.

We wish to thank V. Pasquier and R.A. R\"omer for useful discussions.


\begin{references}

\bibitem{sutherland70} B. Sutherland,  Phys. Rev. A {\bf 4}, 2019 (1970).

\bibitem{sutherland71} B. Sutherland, J. Math. Phys. {\bf 12}, 246 (1971); J. Math. Phys. {\bf 12}, 251 (1971).

\bibitem{jayaprakash84} C. Jayaprakash, C. Rottman and W. F. Saam, Phys. Rev. B {\bf 30}, 6549 (1984).

\bibitem{williams93} E.D. Williams, R.J. Phaneuf, J. Wei, N.C. Bartelt, and T.L. Einstein, Surf. Sci. {\bf 294}, 219 (1993) and  Surf. Sci. {\bf 310}, 251 (1994).

\bibitem{noteHS} The spin-${1 \over 2}$ discrete CS model at $T=0$ was solved by F.D.M. Haldane Phys. Rev. Lett. {\bf 60}, 635 (1988) and B.S. Shastry Phys. Rev. Lett {\bf 60}, 639 (1988) in the isotropic case i.e. for special values of the interaction amplitude ($g=4$) and step density ($p= 1/2$), corresponding to $T_R$ in Fig.~(\ref{diag_phase}). 

\bibitem{herring51} C. Herring, Phys. Rev. {\bf 82}, 87 (1951)

\bibitem{andreev81} A. F. Andreev,
Zh. Eksp. Teor. Fiz. {\bf 80}, 2042 (1981)
[Sov. Phys. JETP {\bf 53}, 1063 (1981)].

\bibitem{heyraud80} J. C. Heyraud and J. J. M\'etois, J. Crystal Growth {\bf 50}, 571 (1980); Acta Met. {\bf 28}, 1789 (1980).

\bibitem{rottman84b} C. Rottman, M. Wortis, J. C. Heyraud and J. J. M\'etois, \prl {\bf 52}, 1009 (1984).

\bibitem{metois87} J. J. M\'etois and J. C. Heyraud, Surf. Sci. {\bf 180}, 647 (1987).

\bibitem{yoon94} M. Yoon, S. G. J. Mochrie, D.M. Zehner, G. M. Watson and D. Gibbs, Phys. Rev. B {\bf 49}, 16702 (1994).

\bibitem{song95} S. Song and S. G. J. Mochrie, \prb {\bf 51}, 10068 (1995)

\bibitem{watson98} G. M. Watson, D. Gibbs, D.M. Zehner, M. Yoon, S. G. J. Mochrie, Surf. Sci. {\bf 407}, 59 (1998).

\bibitem{marchenko81} V.I. Marchenko,
Zh. Eksp. Teor. Fiz. {\bf 81}, 1141 (1981)
[Sov. Phys. JETP {\bf 54}(3), 605 (1981)].

\bibitem{luck83} J. M. Luck, S. Leibler and B. Derrida, J. Physique {\bf 46}, 1135 (1983).

\bibitem{dennijs92} M. den Nijs, Phys. Rev. B {\bf 46}, 10386 (1992).

\bibitem{affleck86} I. Affleck, \prl {\bf 56}, 746 (1986).

\bibitem{villain85} J. Villain, D.R. Grempel and J. Lapujoulade, J. Phys. F {\bf 15}, 809 (1985).

\bibitem{legoff99} E. Le Goff, L. Masson, L. Barbier and  B. Salanon, Surf. Sci. {\bf 432}, 139 (1999).

\bibitem{barbier96} L. Barbier, L. Masson, J. Cousty and B. Salanon, Surf. Sci. {\bf 345}, 197 (1996).

\bibitem{kawakami91} N. Kawakami, \prl {\bf 67}, 2493 (1991).

\bibitem{romer92} R. A. R\"omer and B. Sutherland, Phys. Rev. B {\bf 48}, 6058 (1993).

\bibitem{schultz85} H. J. Schultz, J. Physique {\bf 46} , 257 (1985).

\end{references}
\end{document}